\documentclass[12pt]{amsart}%article}

\usepackage[colorlinks=true]{hyperref}
\usepackage{amsmath,fullpage,color,enumerate}

\usepackage{amssymb}
\usepackage{graphicx}
\usepackage{mathrsfs}
\usepackage{eucal}
\usepackage{bm}
\usepackage{epsfig}
\usepackage{subfigure}
\usepackage{wrapfig}
\usepackage{color}
\usepackage{fancybox}
\usepackage{epstopdf}
\usepackage{color}


\newtheorem{theorem}{Theorem}%[section]
\newtheorem{proposition}[theorem]{Proposition}
\theoremstyle{definition}
\newtheorem{definition}[theorem]{Definition}
\newtheorem{remark}[theorem]{Remark}

\def\R{{\mathbb R}}

\def\cM{{\mathcal M}}
\def\cP{{\mathcal P}}

\def\qhat{\widehat{q}}

\def\thetahat{\widehat{\theta}}
\def\zhat{\widehat{z}}
\def\Qsat{Q_{\rm sat}}
\def\Zad{{Z}_{\rm ad}}
\def\Thad{{\Theta}_{\rm ad}}
\def\Thdisp{{\Theta}_{\rm disp}}
\def\thetaM{\theta^{\rm M}}

\def\rp{\textrm{P}}
\def\ini{\textrm{in}}
\def\fin{\textrm{fi}}

\def\pa{\partial}
\def\be{\begin{equation}}
\def\ee{\end{equation}}
\def\ba{\begin{aligned}}
\def\ea{\end{aligned}}

\newcommand{\DDt}[1]{{D #1\over D  t}}
\newcommand{\pp}[2]{{\pa  #1\over \pa  #2}}

\def\qand{\quad\mbox{and}\quad}

\begin{document}

%\phantom{*}
%\vspace{.15cm}
\underline{\large Study Group Report}
\medskip

\begin{center}
{\Large\bf  Convection in a Single Column}\\[4mm]
{\Large\bf -- Modelling,  Algorithm and Analysis}
\vspace{1.0cm}

\textsf{\begin{tabular}{ll}
  Onno Bokhove & (O.Bokhove@leeds.ac.uk),\\ 
  Bin Cheng & (B.Cheng@surrey.ac.uk), \\
  Andreas Dedner & (A.S.Dedner@warwick.ac.uk), \\
  Gavin Esler & (J.G.Esler@ucl.ac.uk), \\
  John Norbury & (John.Norbury@lincoln.ox.ac.uk),  \\
  Matthew R. Turner & (M.Turner@surrey.ac.uk), \\
  Jacques Vanneste & (J.Vanneste@ed.ac.uk),    \\
  [2mm] \underline{\textrm {Proposer}}: Mike Cullen & (Mike.Cullen@metoffice.gov.uk)
  \end{tabular}
}

\vspace{.6cm}
\bigskip
 
 {\bf ---} \today {\bf ---} 
\end{center}
\vspace{.6cm}
\bigskip

{\bf\large Content}\begin{enumerate}[1.]
\item Introduction.
\item Physical constraints in convection modelling.
\item Settings of the model problem.  
\item Sorting algorithm.
\item Theory of  optimal transport.
\item Conclusion.
\end{enumerate}\bigskip

The group focused on a model problem of idealised moist air convection in a single column of atmosphere. Height, temperature and moisture variables were chosen to simplify the mathematical representation (along the lines of the Boussinesq approximation in  a height variable defined in terms of pressure). This allowed exact simple solutions of the numerical and partial differential equation problems to be found. By examining these, we identify column behaviour, stability issues and explore the feasibility of a  more general solution process.% to be studied.
\bigskip

\section{\bf Introduction}

Atmospheric models used for weather and climate prediction use the classical compressible Navier-Stokes equations, laws of thermodynamics, and laws governing phase changes, radiation and surface fluxes. However, the exact solution of these equations is not computationally feasible, and the equations have to be averaged in space and time before being solved. It is then necessary to design sub-grid models which represent the averaged effects of the unresolved small scales on the resolved solution.

A particularly difficult situation arises with cumulus convection. This is responsible for much of the severe weather outside the tropics, and is the main driver of the tropical circulation which is a fundamental part of the climate system. Cumulus convection can only be directly represented in models with a horizontal grid of at most one or two kilometres. This is not affordable in global climate models at present. The sub-grid modelling of convection is very difficult because of the jump-like  nonlinearity of the moisture-heat exchange processes involved. Such a model has to build in the effects of the large-scale solution  on the convection.

In order to understand these effective large-scale constraints, and thus improve the sub-grid models, it is useful to consider the evolution of a single column of moist air. It is well-known by practising weather forecasters that the occurrence and intensity of convection can be predicted by studying the temperature and moisture profiles in such a column together with knowledge of the external forcing, e.g. see \cite{MetOffice:1993}, chapter 4. The challenge set for the study group was to make a mathematically rigorous version of a single column model which behaved in a way that agreed with the observed behaviour. Such a model can then be used to validate the sub-grid models in daily forecasts, and potentially to improve them. The existence of a rigorous model which describes at least the most important part of the convective process would also indicate high potential deterministic predictability in a situation where deterministic predictions are not yet consistently successful. 

We therefore study the stability of a moist vertical column of air in the presence of external forcing, such as the bodily lifting of the column. Stability requires   that the potential temperature $\theta$ must increase as a function of the height $z$ (in suitably scaled pressure units). But the physics also requires a constraint holds that  the water vapour $q$, i.e. the moisture carried by the air, must be less than a critical level denoted by $\Qsat(\theta,z)$. If not, the excess vapour in an air parcel then condenses and releases heat to the air parcel, thus increasing the potential temperature, and hence possibly changing the stability of the parcel. For simplicity, we treat all condensation as vapour to rain, ignoring things like ice/snow mixtures, and allow the rain to fall out of the column.\bigskip

\section{\bf Physical constraints in convection modelling}
Some key features of the local convection problem are as follows.

$\bullet$ Convective processes happen in a shorter time-scale than that of the horizontal dynamics, and are usually    localized in the horizontal scale.

$\bullet$ Given a parcel, there are   {\bf three constraints} for potential temperature $\theta$,   humidity $q$ and air mass, as the parcel moves vertically. %Here, all motion is vertical, and all variables are functions  only of height $z$ and time $t$.

\indent\indent (i) If the  parcel stays unsaturated, then $\theta,q$ and its air mass are conserved respectively. 

\indent\indent (ii) If the  parcel stays {saturated}, then  its air mass is conserved. The second  constraint simply is $q=\Qsat(\theta,z)$. The third constraint is that the ``moist potential temperature'' $$\thetaM:=\theta+Lq$$ is conserved. Here, $L$ is     a physical constant which gives the latent heat of conversion released to the air parcel when some of the water vapour condenses to precipitation (cloud, rain, ice or some mixture), and $\Qsat(\theta,z)$ is a given function of $\theta,z$ determined by the physical properties of the air and its water vapour.  

%\indent\indent (iii) As the parcel moves, its air mass is conserved.

The conditions (i), (ii) arise from conservation  in both the physical and mathematical senses  of moist thermal energy (measured by the suitably defined potential temperature which is related to the entropy), of moisture as either water vapour or cloud/rain, and of air mass, all defined on the air parcels as they move.

\begin{definition}\label{def:ma}Given the  saturation specific humidity $\Qsat(\theta,z):\,\R^2\mapsto\R$ as a smoothly differentiable function of $\theta,z$ satisfying\be\label{Qsat:mono}
\pp{\Qsat}{\theta}>0\qand\pp{\Qsat}{z}<0,\ee 
the ``moist adiabat'' is a formula associated with  a given parcel (with saturated status) and represented by a curve in the $(\theta,z)$ plane, which obeys the above constraint (ii), i.e. 
\be\label{const:sat}\thetaM=\textsf{constant}=\theta+Lq=\theta+L\Qsat (\theta,z).\ee
The monotonicity conditions \eqref{Qsat:mono} guarantee that there exist two smoothly differentiable functions $\Zad(\cdot,\cdot)$, $\Thad(\cdot,\cdot)$ so that the above moist adiabat  formula is equivalent to\[ z=\Zad(\theta,\thetaM)\;\;\iff \;\;\theta=\Thad(z,\thetaM).\] 

In other words, a moist adiabat curve is the \emph{level set} curve  of function $\theta+L\Qsat (\theta,z)=\thetaM=\textsf{constant}$ in the $(\theta,z)$ plane, which is identical to the  graph of $ z=\Zad(\theta,\thetaM)$, and  of  $\theta=\Thad(z,\thetaM)$, with constant $\thetaM$. 
\end{definition}
Following this definition and applying the chain rule, we obtain:

\begin{proposition}\label{prop:Zad}
 On the moist adiabat with a constant $\thetaM$,
 \[\pp{\Zad}{\theta} ={1+L\pp{\Qsat}{\theta}\over-L\pp{\Qsat}{z}}\qand \pp{\Zad}{\theta} \,\pp{\Thad}{z} =1. \]
\end{proposition}

We can combine this with \eqref{Qsat:mono} to immediately have, with a constant $\thetaM$,
\be\label{ad:mono}\pp{\Thad}{z}>0\qand\pp{\Zad}{\theta}>0.\ee

Now, consider that $\thetaM$ is defined by the parcel's initial configuration and remains constant during the convection process, for both unsaturated and saturated parcels. Then, we can unify these two cases to give a relation between the initial (subscript ``in'') and final (subscript ``fi'') configurations of a given dry/wet parcel experiencing a rising/lowering displacement from $z_\ini$ to $z_\fin$,  
\begin{align}\label{dry:wet}\theta_\fin&=\Thdisp(\theta_\ini,  q_\ini, z_\fin ):=\max\left\{\theta_\ini,\,\Thad(z_\fin,\theta_\ini+L q_\ini) \right\}\,, \\
 L q_\fin&=\theta_\ini+L q_\ini-\theta_\fin.\end{align}
The definition of $\Thdisp$ is regardless of the stability of these configurations, and can be used to describe virtual displacement in the variational formulation. Several remarks are in order. 

1. The use of moist adiabat implies $\thetaM=\theta_\ini+L q_\ini=\theta_\fin+L q_\fin$ for both wet/dry  cases.

2. We   impose the maximum since by \eqref{Qsat:mono} and \eqref{const:sat}, having   $\theta_\fin<\Thad(z_\fin,\thetaM)$  would   mean  \[\ba
\theta_\fin<\Thad(z_\fin,\thetaM)\implies L\Qsat(\theta_\fin,z_\fin)&<L\Qsat\big(\Thad(z_\fin,\thetaM),\,z_\fin\big)\\&=\thetaM-\Thad(z_\fin,\thetaM) <\thetaM-\theta_\fin=L q_\fin\ea\] i.e. $ \Qsat(\theta_\fin,z_\fin)<q_\fin$ which would be unphysical. Interestingly, if we replace all occurrences of $<$ with $=$ above, the calculation still holds; and likewise if we   replace all occurrences of $<$ with $>$ above. Thus, we obtain an \emph{equivalent} saturation condition,
\be\label{sat:equiv} \text{sign}\Big(q- \Qsat(\theta,z)\Big)=\text{sign}  \Big(\Thad(z ,\thetaM)-\theta\Big)\quad\text{where }\;\; \thetaM=\theta+L q.\ee

3. The use of maximum above also means $\theta_\fin\ge\theta_\ini$, namely, only a rising parcel can possibly increase temperature upon condensation of some of its water vapour. Re-evaporation is neglected, and its modelling is possible only if cloud information is included. Our model is time \emph{irreversible}.

\subsection*{Convective instability and monotonicity of  $\Qsat$}Convective instability is tied to      the key properties that $\Qsat$    increases with $\theta$  and decreases with height $z$ (c.f. \eqref{Qsat:mono}). First, the column is unstable whenever $\theta$ is not monotonically increasing in $z$. Further, even the monotonicity holds, we still need to check the saturation condition.
When the actual humidity of a parcel reaches the saturation level $q=\Qsat$, condensation takes place and latent heat is released, resulting in increase in $\theta$. In turn, the parcel is lifted up by buoyancy. Then, one can use the monotonicity of $\Qsat$ in \eqref{Qsat:mono} and Proposition \ref{prop:Zad} to show   that following any  {rising, saturated  parcel}, the Lagrangian derivative $\DDt\Qsat  <0$. This decrease in $\Qsat$ will encourge more condensation and thus release of more latent heat, resulting in  a positive feedback mechanism for a saturated parcel to rise.  

For the parcel to \emph{continue} rising (i.e. convective instability) however, we need more. If $\theta$ is differentiable in $z$, we propose two conditions related to  convective instability,  
\begin{align}\label{cond:start}\text{``triggering condition''}:\qquad \pp{\Thad}{z}(z,\thetaM) \ge\pp{\theta}{z}&\qand q=\Qsat(\theta,z),\\
 \label{cond:stop}\text{``stopping condition''}:\qquad  \pp{\Thad}{z}(z,\thetaM) <\pp{\theta}{z}&,\end{align}
the latter of which defines  the bottom of an ``inversion layer''  or  is near the tropopause.
Note that the inversion layer can be dynamically changing with time and actually dependent on the $\thetaM$ of each parcel. 

These two conditions are consistent with the maximisation of $\int z\theta \, dz$ over all rearrangements of the parcels, i.e. measure-preserving maps, which is carried out in the algorithm below. The conditions can be shown using a variational argument with the perturbation being swapping of two small measures/blocks of parcels. In the general situation where $\theta$ is not necessarily differentiable, the conditions \eqref{cond:start}, \eqref{cond:stop} are applied on small blocks of air at different heights (locally or globally) by comparing their temperatures.\bigskip

\section{\bf Settings of the model problem}

   Consider a single vertical column that is horizontally uniform. It is being uniformly lifted on a longer timescale than that of the convective adjustment.% ({\clrr??? 1 degree = 60 km/h}) 
   
   \underline{Assumption  1}. The column   responds/makes adjustment, according to the physical environment's thermodynamic  change, but any feedback to the environment is neglected.

\underline{Assumption  2}. For simplicity, physical processes such as background thermal radiation and  ice/water re-evaporation are neglected so that a given parcel's humidity $q$ never increases, and so here  its $\theta$ never decreases.

\underline{Assumption  3}. There are no mass/heat/water vapor fluxes at the top and bottom boundaries. 

The column may be rising as a whole at a given speed and still satisfying the no flux boundary conditions. This is a simplified model for a cold front wedging at a constant rate under a column of  moist air.  The wedging is at a slower rate than the convective adjustment. 

By using a co-moving frame, the spatial domain is fixed as $z\in[0,1]$. Then,  the term $(-\alpha t)$ in \eqref{Qsat:exp} accounts for the column's actual upward motion.

Under the hydrostatic assumption, for the single column model, the actual height and pressure are monotonically linked. Then, we use Hoskins' pseudo-height $z$,
\[z:=\Big\{1-\big({p\over p_0}\big)^{R/c_p}\Big\}z_a\]so that the (environmental) pressure variable becomes implicit. The constant   $z_a>1$.

For simplicity and to provide explicit examples, we adopt an accurate approximation (\cite{Lock:Norbury})
\be\label{Qsat:exp} 
\Qsat(\theta,z)=A_0\exp\left(r\left[\theta-\beta z-\theta_{\rm PBL}-\alpha t\right]\right),
\ee
where $A_0,~r,~\beta,~\theta_{\rm PBL}$ and $\alpha$ are constants. The $\Zad$ function of Definition \ref{def:ma} is then explicitly given as
\be\label{Zad:exp} 
\Zad(\theta,\thetaM)={1\over\beta}\left[\theta-{1\over r}\ln{\thetaM-\theta\over LA_0} -\theta_{\rm PBL}-\alpha t\right].
\ee
 In light of Proposition \ref{prop:Zad}, we have that on the moist adiabat with constant $\thetaM$,
\be\label{Zad:th:exp} \pp{\Zad}{\theta} ={1\over\beta}\left[1+{1\over r\,( \thetaM-\theta)}\right]={1\over\beta}\left[1+{1\over r\,L\,\Qsat}\right].\ee
Therefore, the threshold  used in instability/stability conditions \eqref{cond:start},   \eqref{cond:stop} is explicitly
\be\label{Thad:z:exp}\pp{\Thad}{z}(z,\thetaM)={ \beta}\left[1+{1\over r\,L\,\Qsat}\right]^{-1}.\ee

\subsection*{Rearrangement/adjustment problem}
Based on Lock and Norbury \cite{Lock:Norbury} and Goldman's MSc thesis \cite{Goldman}, we make

\underline{Assumption 4}. The column   responds/makes adjustment at a much shorter time scale compared to the environment's thermodynamic change. So we do \emph{not} model the actual  {dynamics}, i.e. the acceleration and deceleration of a saturated parcel rising due to buoyancy. Mathematically, the response time is infinitesimal.

{\bf Main idea}: the potential energy $\int z\theta \,dz$ is maximised in a way consistent with the moist adiabat. See the next two sections.

\begin{remark}
The   PDEs (7), (8) in Goldman's thesis \cite{Goldman} are basically the same as  Definition \ref{def:ma} of  ``moist adiabat''. That result was regarding the existence of weak solutions to (7), (8) in \cite{Goldman} under some stability condition -- no uniqueness was proven. The maximisation of   $\int z\theta \,dz$ was not explicitly stated in the theorems there, but in the construction of weak solutions, the proof of \cite{Goldman} uses a rearrangement strategy to enforce a  certain monotonicity condition which, combined with his stability condition, \emph{may} just maximize $\int z\theta \,dz$. Goldman claims his version of the stability condition is consistent with that of Cullen \& Purser \cite{Cullen:Purser}.
\end{remark}
\bigskip

\section{\bf Sorting algorithm}

Recall constraints (i),  (ii) and Definition \ref{def:ma} of  ``moist adiabat''.

Dependent variables are $\theta_\rp,q_\rp,z_\rp$ with subscript P indicating they are Lagrangian variables on a parcel. These variables are subject to either the  unsaturated constraint, i.e. $\theta_\rp,q_\rp$ remain constant, or the saturated constraint, i.e. $q_\rp=\Qsat(\theta_\rp,z_\rp)$ and   $\theta_\rp=\Thad(z_\rp,\thetaM_\rp)$. Mass is conserved due to the fact that the algorithm simply rearranges the discrete blocks of parcels.
 
 The saturation function $\Qsat$ is given by \eqref{Qsat:exp}. For now consider $t=0$.

The numerical sorting algorithm assumes that the atmosphere is divided into $N$ parcels with initial temperatures  $\theta_1^0,~\theta_2^0,...,\theta_N^0$ and moistures $q_1^0,~q_2^0,...,q_N^0$ (some of which are at saturation) at the heights $z_1,~z_2,...,z_N$. For each parcel the combination $\theta_i^M=\thetahat_i+L\qhat_i=\theta_i^0+L q_i^0$ is conserved during the rearrangement.  Here the hat on the variables denotes the value of $\theta_\rp$, $q_\rp$ and $z_\rp$ during the sort.   Thus if the moisture level changes due to condensation, then the temperature increases. With the   function defined in \eqref{dry:wet} that unifies the dry and wet cases, we can write
\[\thetahat_i=\Thdisp(\theta_i^0, q_i^0,\zhat_i)\qand \qhat_i= (\theta_i^0+Lq_i^0-\thetahat_i)/L.\]

The algorithm works by starting at height $z_N$. It then temporarily lifts every parcel below this height up to $z=z_N$ with the would-be $\thetahat_i$ determined by the above formula.
Once this is done for all parcels below $z_N$,   the parcel with the largest temperature (e.g. labelled $m$) is chosen to be at height $z_N$. Parcel $m$ is then eliminated from the sort.   Parcels previously labelled  $(m+1)$ to $N$ have their heights lowered by one grid. 

The code then moves to $z_{N-1}$ and repeats until it reaches $z_1$.

\begin{figure}
\begin{center}
\includegraphics[width=0.95\textwidth]{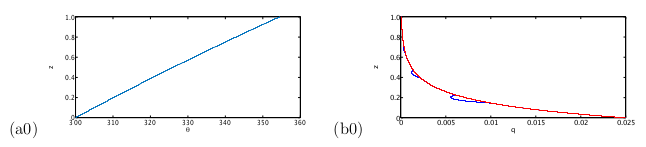}
\end{center}
\caption{Plot of (a0) is the initial temperature distribution $\theta^0_i(z_i)$ and (b0) is the initial moisture distribution $q^0_i(z_i)$ (blue line) for $N=5000$ parcels. In Plot (b0) the red line denotes $\Qsat(\theta_i^0,z_i)$.}
\label{fig:ics}\bigskip\bigskip\bigskip\hrule\bigskip\bigskip\bigskip
\end{figure}

\begin{figure}
\begin{center}
\includegraphics[width=0.95\textwidth]{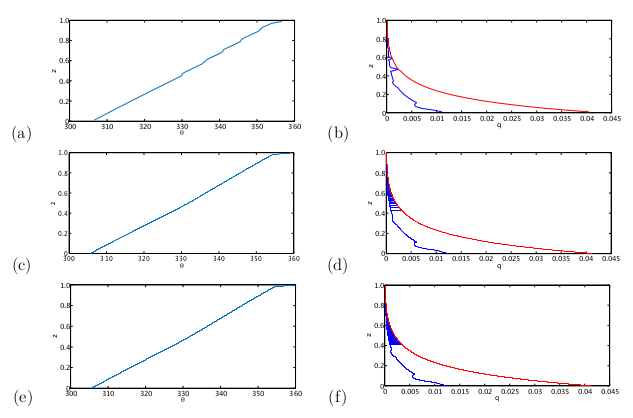}
\end{center}
\caption{Plots of (a), (c), (e) are the final temperature distribution $\theta_i(z_i)$ and (b), (d), (f) are the final moisture distribution $q_i(z_i)$ where the red curves   denote the new $\Qsat$. 
The number of parcels is:   $N=50$ in  (a), (b);    $N=500$ in (c), (d);    and   $N=5000$ in (e), (f).  }
\label{fig:results}
\end{figure}

Results of this code are presented in Figure \ref{fig:ics} for the initial configuration and in Figure \ref{fig:results} for the final configuration. We choose the following parameters  
\[
  z_i=\frac{i-0.5}{N}\qquad i=1,...,N,
\]
and initial temperature configuration 
\[
\theta_i^0= 300\exp\left(\frac{z_i}{6 }\right),
\]
where
\[
A_0=0.025,~~~r=0.09,~~~\beta=120,~~~\theta_{\rm PBL}=300,
\]
in $\Qsat$. The value of $L=2400$. The initial moisture levels are taken to be% (rescale to $z_i\in[0,1]$)
\[
   q_i^0=\min \Big \{1,\, \frac{5}{4}+\frac{1}{2}\sin({8\pi z_i}) \Big \}\,\Qsat(\theta_i^0,z_i)\,.
\]

The results in Figure \ref{fig:results} show that the numerical sorting algorithm appears to converge as $N$ is increased. At this stage the visual convergence is sufficient, but more detailed numerical analysis is required to confirm this. As $N$ increases the system moves towards that of unsaturated air for   $z\lesssim 0.4$ and approximately a region of saturated air parcels above this.   The highly oscillatory behaviour of moisture in Plot (f) may suggest that as $N\to\infty$, the algorithm converges to some kind of transport \emph{plan} rather than a transport \emph{map}. Also note that the final temperature distribution contains step jumps in the temperature for $N=50$ and $500$, but these get smoothed out as the number of parcels increases to $N=5000$.  However, the final temperature distribution and its smoothness depend on the given initial data.

\section{\bf Theory of  optimal transport}

For SG motion in the $x-z$ plane (\cite{Holt}), a saturated parcel moves along an adiabatic surface and conserves the $x$- momentum (since there is no movement in the $y$-direction). If the energy minimisation property in the dry case also applies here, then the total energy $\big(-\int z\theta\,dz  + $kinetic energy$\big)$  is minimised. But since the kinetic energy is constant, we must have that $\int z \theta \,dz$ is maximized.

Back to the column problem.
For the transport map, it is represented by a function $\sigma: [0,1]\mapsto[0,1]$ which preserves the Lebesgue measure, i.e., for any $0\le a\le b\le1$,
\[\int_a^b\sigma(z)\,dz=b-a.\]
Let $\cM$ be the collection of all such measure preserving {\it maps}. Recall $\Thdisp$ defined in   \eqref{dry:wet}. Then we look for (maybe more than one)   maximizer
\[\sigma^*=\arg \max_{\sigma\in\cM}\Big\{\int_0^1 \sigma(z)\, \Thdisp(\theta(z),q(z),\sigma(z))\,dz \Big\}.\]
%Recall the monotonicity of  $\Thad(\cdot,\cdot)$. So,   $\max\big\{\Thad(\sigma(z),\theta(z)+Lq(z)),\,\theta(z)\big\}$ will choose the first argument if the parcel at $z$ is saturated and rises i.e. $\sigma(z)> z$, and it will choose the second argument if  the parcel at $z$ stays or sinks, i.e. $\sigma(z)\le z$. The condition ``$\sigma(z)\le z\text{\;\; or\;\; }q(z)=\Qsat(\theta(z),z)$'' makes sure only saturated parcel rises -- although being saturated does not guarantee a higher spot!

For the transport plan, it is represented by a (generalized) nonnegative function $K(z,z')$ defined on $ [0,1]\times[0,1]$. It quantifies the percentage of parcels moved from $z$ to $z'$. It  preserves things in the sense of marginal probability measures,
\[\int_0^1K(z,z')\,dz=\int_0^1K(z,z')\,dz'=1.\]
The transport map $\sigma(z)$ is a special case $K(z,z')=\delta(z'-\sigma(z))$. So, in general, $K(\cdot,\cdot)$ is a   measure (nonnegative distribution) defined on $ [0,1]\times[0,1]$.

Let $\cP$ be the collection of all such measure preserving {\it plans}. %Recall $\Thad$ defined in Proposition \ref{prop:Thsat}. 
Then we look for (maybe more than one)   maximizer
\[K^*=\arg \max_{K\in\cP}\Big\{\int_0^1\int_0^1 z'\,K(z,z')\,\Thdisp(\theta(z),q(z),z')\,dz\,dz' \Big\}.\]
\bigskip

\section{\bf Conclusion}

We constructed a numerical algorithm whose solutions converged to exact solutions  that we found for some idealised test cases. But we also found that the numerical solutions for certain initial moisture and temperature data appeared to converge to {\it``new''} types of solutions which correspond to transport plans (rather than maps) for the variational problems. These new solutions are extremely challenging for conventional fluid dynamics codes because they are highly, and discontinuously, oscillatory and vary rapidly from wetter to drier layers. The sorting algorithm appeared to find that different solutions may exist for the same initial data and this suggests further work in how to devise computer code in practical forecasting.

%Recall the monotonicity of  $\Thad(\cdot,\cdot)$. So, $\max\big\{\Thad(z',\theta(z)+Lq(z)),\,\theta(z)\big\}$ will choose the first argument if $K(z,z')$ fraction of the parcel at $z$ is saturated and rises to $z' > z$, and it will choose the second argument if  $K(z,z')$ fraction the parcel at $z$ stays or sinks to $z'\le z$. This allows one to split the above double integral into two parts, one for $z'>z$ and one for $z'\le z$, without using the $\max$ function. The condition ``$\text{supp}K(z,\cdot)\subset[0,z]\text{\;\; or\;\; }q(z)=\Qsat(\theta(z),z)$'' makes sure only saturated parcel rises -- although being saturated does not guarantee a higher spot!

%\bibliographystyle{}

%\bibliography{../sloshbib}

\bigskip

\begin{thebibliography}{10}
\bibitem{Cullen:Purser} Cullen, M. J. P., and R. J. Purser. ``An extended Lagrangian theory of semi-geostrophic frontogenesis." Journal of the atmospheric sciences 41, no. 9 (1984): 1477-1497.
\bibitem{Goldman} Goldman, Dorian. ``Weak Lagrangian Solutions to a One Dimensional Model of the Moist Semi-geostrophic Equations." M.Sc. diss., University of Toronto, 2008.
\bibitem{Holt} Holt, M. W. ``Semigeostrophic moist frontogenesis in a Lagrangian model." Dynamics of Atmospheres and Oceans 14 (1989): 463-481.
\bibitem{Lock:Norbury} Lock, A. M., and J. Norbury. ``A column model of moist convection: some exact equilibrium solutions." Quarterly Journal of the Royal Meteorological Society 137, no. 657 (2011): 979-991.
\bibitem{MetOffice:1993} Met Office {``Forecaster's Reference Book''}, 2nd ed. Met O. 1012 (1993).
\end{thebibliography}
\newpage

\end{document}